\documentclass[aps,prl,twocolumn,superscriptaddress]{revtex4-1}  
\usepackage{graphicx}  
\graphicspath{{figs/}}
\usepackage{dcolumn}   
\usepackage{bm}        
\usepackage{amssymb}   
\usepackage{amsmath}   
\usepackage{amsthm}    
\usepackage{xcolor}    
\usepackage{colortbl} 
\usepackage{tikz}      
\usepackage{ifthen}    
\usepackage{multirow}  
\usepackage{tabu}      
\usepackage{soul}      
\usepackage{subfigure}
\usepackage{xspace}
\usepackage{braket}
\usepackage[unicode=true,
  linktocpage,
  linkbordercolor={0.5 0.5 1},
  citebordercolor={0.5 1 0.5},
  linkcolor=blue]{hyperref}
\usepackage{hyperref}
\hypersetup{colorlinks=false, citecolor=blue, urlcolor=blue, linkcolor=blue}
\usepackage[utf8]{inputenc}
\usepackage[english]{babel}
\usepackage{natbib}

\newcounter{para}

\begin{document}

\title{Observation of 4- and 6-magnon bound-states in the spin-anisotropic frustrated antiferromagnet FeI$_2$}

\author{Ana\"elle Legros}
	\thanks{These authors contributed equally to this work}
	\affiliation{The Institute for Quantum Matter and the Department of Physics and Astronomy, The Johns Hopkins University, Baltimore, MD 21218, USA}
\author{Shang-Shun Zhang}
	\thanks{These authors contributed equally to this work}
	\affiliation{Department of Physics and Astronomy, University of Tennessee, Knoxville, TN 37996, USA}
\author{Xiaojian Bai}
	\affiliation{School of Physics, Georgia Institute of Technology, Atlanta, GA 30332, USA}
	\affiliation{Neutron Scattering Division, Oak Ridge National Laboratory, Oak Ridge, TN 37831, USA}
\author{Hao Zhang}
	\affiliation{Department of Physics and Astronomy, University of Tennessee, Knoxville, TN 37996, USA}
	\affiliation{Materials Science and Technology Division, Oak Ridge National Laboratory, Oak Ridge, TN 37831, USA}
\author{Zhiling Dun}
	\affiliation{School of Physics, Georgia Institute of Technology, Atlanta, GA 30332, USA}
\author{W. Adam Phelan}
	\affiliation{PARADIM, Department of Chemistry, The Johns Hopkins University, Baltimore, MD 21218, USA}
\author{Cristian D. Batista}
	\affiliation{Department of Physics and Astronomy, University of Tennessee, Knoxville, TN 37996, USA}
\author{Martin Mourigal}
	\affiliation{School of Physics, Georgia Institute of Technology, Atlanta, GA 30332, USA}
\author{N. P. Armitage}
	\email[To whom correspondence should be addressed, ]{npa@jhu.edu}
	\affiliation{The Institute for Quantum Matter and the Department of Physics and Astronomy, The Johns Hopkins University, Baltimore, MD 21218, USA}
\maketitle


\textbf{Spin-waves {\it e.g.} magnons are the conventional elementary excitations of ordered magnets. However, other possibilities exist. For instance, magnon bound-states can arise due to attractive magnon-magnon interactions and drastically impact the static and dynamic properties of materials~\cite{torrance_excitation_1969, hoogerbeets_evidence_1984}.  Here, we demonstrate a zoo of distinct multi-magnon quasiparticles in the frustrated spin-1 triangular antiferromagnet FeI$_2$ using time-domain terahertz spectroscopy.  The energy-magnetic field excitation spectrum contains signatures of one-, two-, four- and six-magnon bound-states, which we analyze using an exact diagonalization approach for a dilute gas of interacting magnons.  The two-magnon single-ion bound states occur due to strong anisotropy and the preponderance of even higher order excitations arises from the tendency of the single-ion bound states to themselves form bound states due to their very flat dispersion.  This menagerie of tunable interacting quasiparticles provides a unique platform in a condensed matter setting that is reminiscent of the few-body quantum phenomena central to cold-atom, nuclear, and particle physics experiments.}


Insulating spin systems are widely studied for their unconventional ground states, exotic magnetic excitations and spin textures. Geometrically frustrated lattices, like the 2D antiferromagnetic triangular lattice, can lead to the suppression of magnetic order at low temperatures and the emergence of fractionalized excitations~\cite{broholm_quantum_2020, gingras_quantum_2014}. But even a long-range ordered magnetic phase can exhibit excitations that are entirely distinct from conventional magnons. For instance, depending on spin-space anisotropies and the range of  magnetic interactions, single-magnon quasiparticles can interact attractively with each other, generating multi-magnon bound states. The existence of such bound states was first predicted in the 1930s by Bethe~\cite{bethe1931theorie} in 1D quantum magnets, using a spin-1/2 Heisenberg model. These exchange-driven bound-states are usually observed in 1D ferromagnetic spin chains with \textit{S} $\leq$ 1~\cite{torrance_excitation_1969, hoogerbeets_evidence_1984, chauhan_tunable_2020}. Although two-magnon bound states have been detected in experiments, evidence for even higher order magnon bound states remains scarce. Recently, a 3-magnon bound state was observed in a quasi-1D antiferromagnetic spin chain system with $S\!=\!2$~\cite{dally_three-magnon_2020}.
Because of their potential for a deeper understanding of fundamental phenomena in magnetism and of few-body problems (\textit{i.e.} quantum mechanics of a finite number \textit{n}$>$2 of interacting particles), such excitations are of intense current interest in the study of insulating magnets~\cite{Nishida_2013,Kato_2020,wang2018experimental, keselman2020dynamical, yoshida2017spin, pradhan2020two, ward2017bound, wulferding_magnon_2020}. But they are also important for potential technological applications since they can for example strongly affect the transport properties in a one-dimensional chain of qubits \cite{subrahmanyam_entanglement_2004}. 

\begin{figure*}[t]
\includegraphics[width=1.\textwidth]{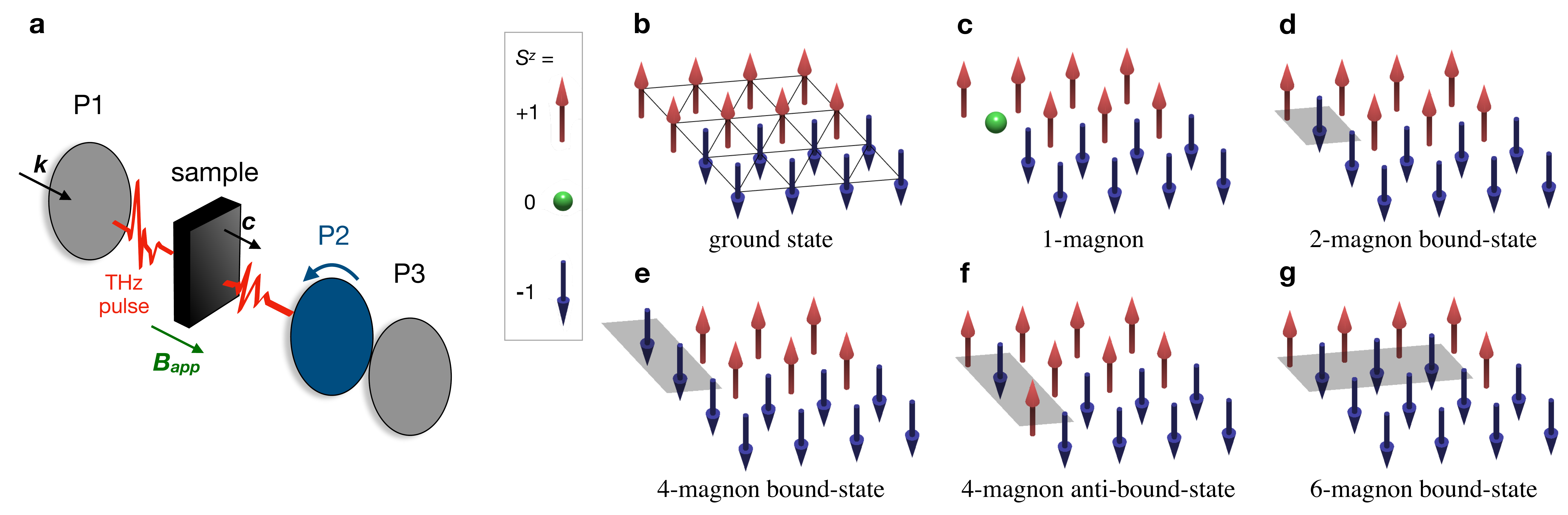}%
\caption{\textbf{Time-domain terahertz spectroscopy set-up and multimagnon excitations of FeI$_2$.}
\textbf{a.} Sketch of the experimental time-domain terahertz spectroscopy set-up~\cite{morris_polarization_2012}: a polarizer (P1) linearly polarizes a terahertz waveform that is transmitted through the sample; the sample induces ellipticity; the fast rotating polarizer P2 modulates the polarization. A DC magnetic field is applied along the crystallographic \textit{c}-axis of the sample in the Faraday geometry (the magnetic field is parallel to the THz propagation). \textbf{b-g.} Single-ion states of Fe$^{2+}$ ions accesible at low-energy correspond to $S\!=\!1$ magnetic moments with uniaxial anisotropy. The sketches represent a plane of spins in the \textbf{b.} ground state configuration, and with examples of \textbf{c.} a one-magnon elementary excitation (spin wave), \textbf{d.} a two-magnon bound state (SIBS), \textbf{e.} a 4-magnon bound state, \textbf{f.} a 4-magnon anti-bound state, and \textbf{g.} a 6-magnon bound state. }
\label{Fig.Experiment}%
\end{figure*}

Here we study multi-magnon bound state excitations in the spin-anisotropic triangular lattice antiferromagnet FeI$_2$. The magnetic Fe$^{2+}$ ions in this compound carry $S\!=\!1$ and are distributed on hexagonal planes (Fig.~\ref{Fig.Experiment}). In zero magnetic field, FeI$_2$ orders spontaneously in a striped antiferromagnetic phase below $T_N$ $\approx$ 9~K~\cite{fert_phase_1973}. One distinctive feature for the $S\!=\!1$ spins in FeI$_2$ is that the energy of two spin deviations on a single-site -- a particular form of 2-magnon excitation called single-ion bound-state (SIBS, Fig.~\ref{Fig.Experiment}{\bf d}) -- is comparable or even lower than the energy of a single magnon (Fig.~\ref{Fig.Experiment}{\bf c}). Recent neutron scattering experiments in zero magnetic field at \textit{T} $<$ $T_N$ revealed that this effect stems primarily from the balance between strong easy-axis single-ion anisotropy $D$ and nearest-neighbor ferromagnetic exchange interaction $J_1$ \cite{bai_hybridized_2020}, although competing further-neighbor exchange interactions are necessary to explain the complex magnetic structure and details of the excitation spectrum. Modeling of these neutron scattering measurements also evinced that the large spectral weight and weak dispersion of the SIBS originates from a hybridization with the single-magnon band through off-diagonal exchange interactions. As we will see below, the almost flat band nature of the SIBS promotes further bound states formation and allows for a very rich phenomenology to develop.

In the presence of a $\boldsymbol{c}$-axis magnetic field, FeI$_2$ undergoes several metamagnetic transitions~\cite{wiedenmann_neutron_1988}; we only study the low-field antiferromagnetic phase here (at \textit{T}~$=$~4~K and $\mu_0$\textit{H}~$<$~5~T). Using time-domain THz spectroscopy (TDTS) we find evidence for a variety of low energy multi-magnon excitations (up to 6-magnon character), along with interactions and hybridization between them. Through a comparison with exact diagonalization calculations for a generalized spin-wave Hamiltonian, our work elucidates how hybridization and interactions between magnetic excitations with different quantum numbers stabilize a low-energy subspace with at least 4 distinct types of quasiparticles. The presence of distinct low-energy excitations that can be tuned by magnetic field is of general interest for the comprehensive understanding of interacting quasiparticles in materials.  

We conducted TDTS measurements as a function of magnetic field ($\mu_0H\leq5$~T along the crystal $\boldsymbol{c}$ axis) to study the low-energy magnetic excitations of FeI$_2$ ($\nu \leq 1.5$~THz $\equiv 6.2$~meV) in the antiferromagnetic phase at $T\!=\!4$~K. The Faraday geometry of the experiment (Fig.~\ref{Fig.Experiment}{\bf a}) allows extracting the transmission eigenstates for both left-handed (LCP) and right-handed circular polarization (RCP) via the polarization modulation technique (see Methods). Employing long time scans of $50$~ps along with reference spectra at higher temperatures \cite{morris_hierarchy_2014} yields a high-frequency resolution appropriate for distinguishing resonance peaks separated by $\Delta\nu\approx0.02$~THz (0.08 meV). While this procedure allows measuring fine features in frequency-dependent spectra, it does not allow us to extract the magnitude of the transmission coefficients with quantitative precision. We thus normalize intensity plots to their maximum value. 

\begin{figure*}[t]
\includegraphics[width=2.0\columnwidth]{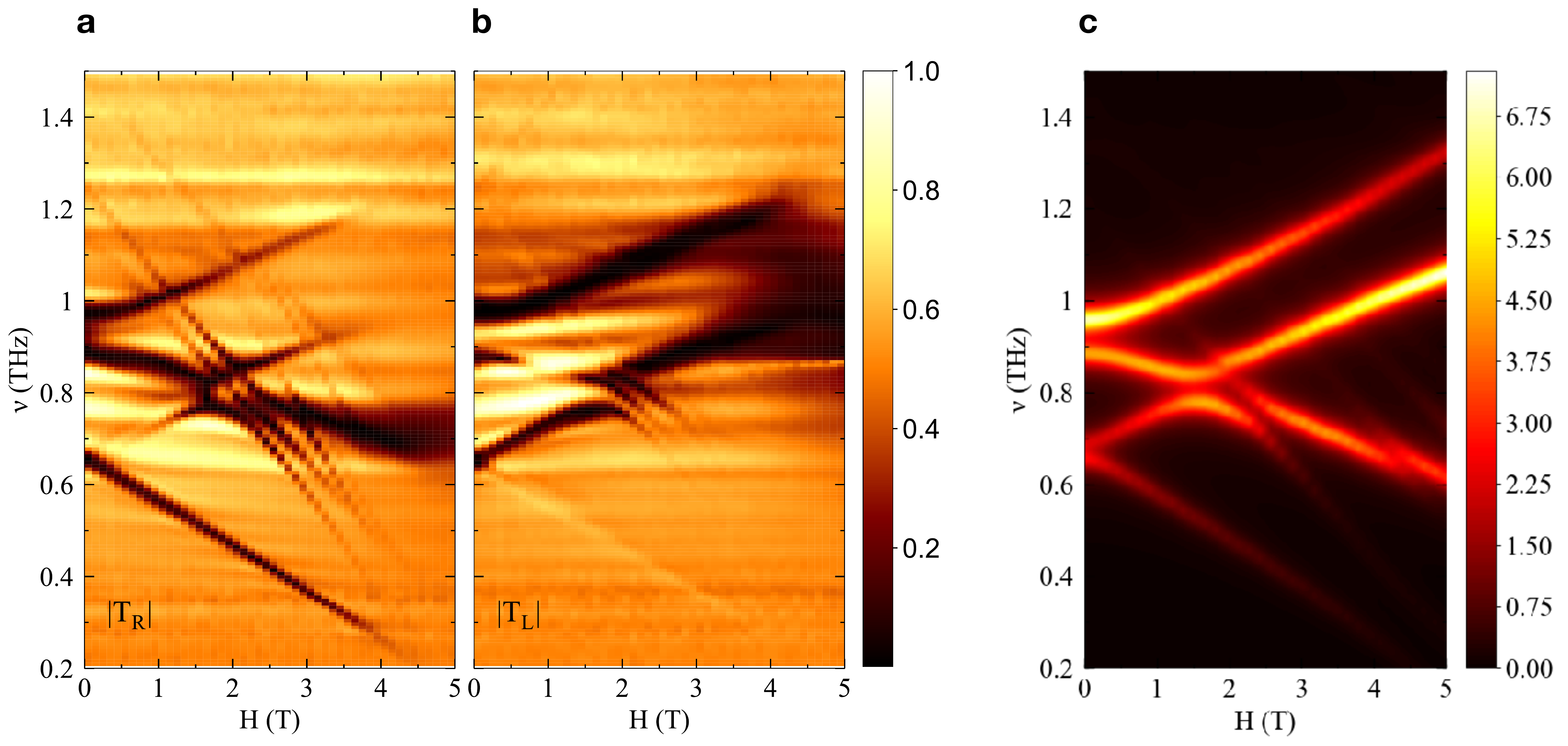}
\caption{\textbf{Magneto-transmission spectra below the Néel temperature.}
Magnitude of the transmission coefficients for \textbf{a.} right- and \textbf{b.} left-handed circularly polarized lights (normalized to 1 for the maximum value), at \textit{T} = 4K, as a function of frequency and magnetic field. Horizontal striations on the plots are experimental artifacts. \textbf{c.} Calculated $\omega[ {\chi''_{xx}} $ (\textit{q} = 0, $\omega$) + $\chi''_{yy}$(\textit{q} = 0, $\omega$)] as a function of energy (or associated frequency) and magnetic field, using the exact diagonalization Lanczos method and the model parameters described in the SI.}
\label{Fig.TR_TL}%
\end{figure*}

The frequency ($\nu$) and magnetic-field ($\mu_0H$) dependence of the complex transmission functions, plotted in magnitude as \textit{$|T_R|$} (RCP, Fig.~\ref{Fig.TR_TL}{\bf a}) and \textit{$|T_L|$} (LCP, Fig.~\ref{Fig.TR_TL}{\bf b}) shows dark regions corresponding to strong absorption. We associate these absorption lines with a wealth of distinct magnetic excitations with several crossings and apparent hybridizations. At fixed frequency and magnetic field, most branches show a different response in the RCP and LCP channels; several resonances are present in both channels. The steep field-slope of several branches is clear evidence for their multi-magnon character, as discussed further below. Earlier far-infrared spectroscopy and electron spin resonance studies~\cite{petitgrand_magnetic_1980, katsumata_single-ion_2000} observed the zero-field absorption branches at $\nu = 0.65, 0.88$ and $0.97$~THz. The high energy and field resolution of our experiments uncover new and very steep branches,
as well as additional details that are important to fully understand the low-energy excitations of FeI$_2$ (see Supplemental Information for comparison to earlier results).

\begin{figure*}[t]
\includegraphics[width=2.0\columnwidth]{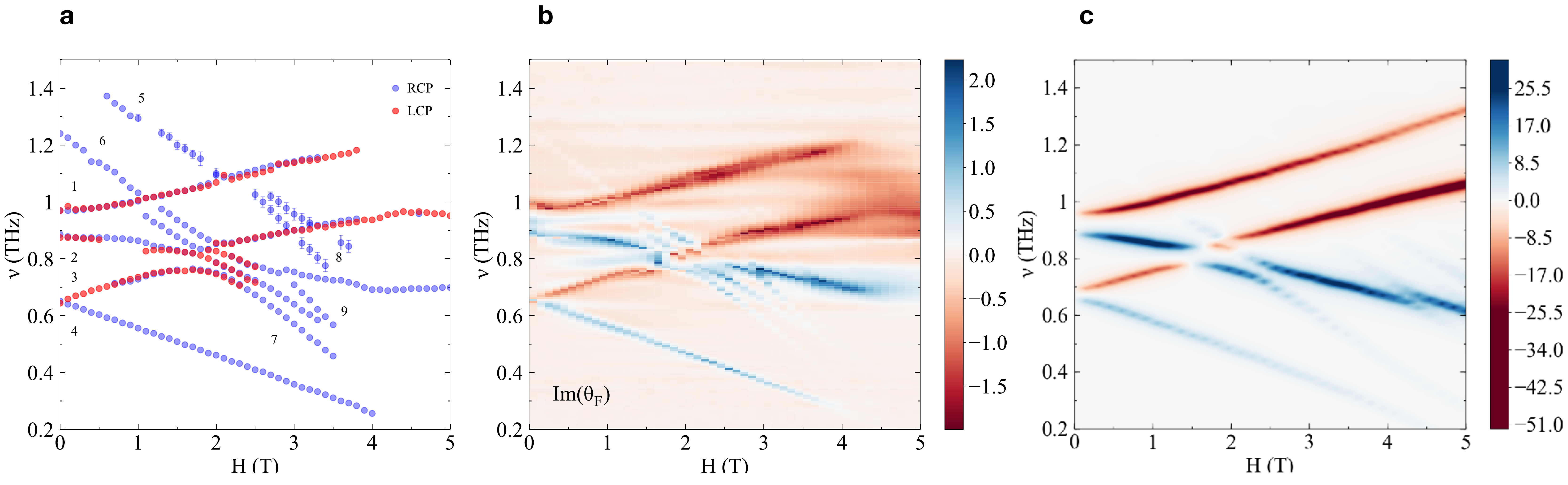}
\caption{\textbf{Excitation branches and spectrum ellipticity.}
\textbf{a.}~Position of the magnetic resonances at \textit{T} = 4K as a function of frequency and magnetic field.  All resonances in both RCP and LCP channels are represented (with some overlap). All resonances correspond to dips in the transmission spectra, except for the data points with error bars which are only distinguishable in the 2D color plot of Fig. 2{\bf a} and assigned by eye. The numbers are used to label the different excitation branches. \textbf{b.}~Imaginary part of the complex Faraday angle $\theta_F$ (ellipticity) in arbitrary units, at \textit{T} = 4~K, as a function of frequency and magnetic field.  \textbf{c.}~Calculated  $\omega\left[\chi''_{+-}(q=0, \omega) - \chi''_{-+}(q=0, \omega)\right]$ as a function of energy (frequency) and magnetic field, using the exact diagonalization (ED) Lanczos method. The model parameters are detailed in the SI.}%
\label{Fig.Ellipticity}%
\end{figure*}

To gain further insight, we extract the absorption lines' position by inspecting the spectra for both polarization channels (Fig.~\ref{Fig.Ellipticity}{\bf a}). Weak modes are sometimes difficult to track from these spectra, and we infer their energies from 2D intensity plots, along with error-bar estimates. In addition to elucidating the slope of all observed branches, \textit{i.e.} the different $\boldsymbol{c}$-axis magnetization $S^z$ of the underlying excitations, this data elucidates crossing and hybridization between modes with different slopes. Given that the uniform magnetization is small below $5$~T, these excitations can be further characterized via the sign of $S^z$ relative to the magnetic field. In Faraday geometry, this is related to the circular dichroism of the material, which we plot using the imaginary part of the complex angle $\theta_F = \arctan[(T_R-T_L)/i(T_R+T_L)]$, the ellipticity (see Methods), as a function of frequency and magnetic field (Fig.~\ref{Fig.Ellipticity}{\bf b}). As absorptions mainly appear in either the RCP or LCP channel, this accentuates the fact that different excitation branches exhibit different signs of $S^z$ (Fig.~\ref{Fig.Ellipticity}{\bf b}).

The evolution of the excitation energies with the magnetic field yields effective \textit{g}-factors (see SI for extracted values) from which we identify at least four different kinds of excitations, including the previously reported single modes and 2-magnon excitations (branches labeled \#1 to \#4 in Fig.~3\textbf{a}), the latter with single-ion bound-state (SIBS) character (Fig.~\ref{Fig.Experiment}{\bf c}--{\bf d}). Given that the SIBS excitations have an almost flat dispersion throughout the Brillouin Zone~\cite{bai_hybridized_2020}, and given the ferromagnetic nature of nearest-neighbor exchange interactions, SIBS display a strong propensity to form bound states with themselves. As a result, we infer that excitations with the highest effective \textit{g}-factors (branches labeled \#5 to \#9 in Fig.~3\textbf{a}) are exchange bound states comprising primarily two and three SIBS, \textit{i.e.} 4-magnon (Fig.~\ref{Fig.Experiment}{\bf e}) and 6-magnon bound states (Fig.~\ref{Fig.Experiment}{\bf g}), respectively. For all excitations, we observe effective \textit{g}-factors that do not reach precisely the maximum hypothetical values of twice, four times, or six times that of single magnon excitations. We associate this effect to the hybridization between the different excitations, {\it e.g.} a 4-magnon bound state can still be regarded as a bound state of two SIBS, but other states mix in when proximate in energy.  In fact, similar to neutron scattering experiments~\cite{bai_hybridized_2020}, the multi-magnon bound states in FeI$_2$ only become detectable in TDTS due to their hybridization with single magnon modes.

Additional insight on the role of multi-magnon hybridization in FeI$_2$ is gained by calculating the absorption spectra using the microscopic spin-exchange Hamiltonian
\begin{eqnarray}
{\cal H}=\sum_{\langle ij\rangle}\sum_{\mu\nu}{S}_{i}^{\mu}{\cal J}_{ij}^{\mu\nu}{S}_{j}^{\nu}-\sum_{i}\left[ D Q_{i}^{zz}+\mu_{0}\mu_{B}gHS_{i}^{z}\right], \, \label{eq:hamiltnian}
\end{eqnarray}
where $Q_i^{zz}=(S_i^z)^2 - 2/3$, the values of the exchange parameters ${\cal J}_{ij}^{\mu \nu}$ and single-ion anisotropy $D$ have been established from the neutron scattering data~\cite{bai_hybridized_2020}, and $g=3.2$~\cite{petitgrand_magnetic_1980}. As previously demonstrated, the nearest-neighbor interactions in FeI$_2$ are spatially-anisotropic and include symmetric off-diagonal terms, $J_1^{z \pm}$ and $J_1^{\pm \pm}$, that are responsible for the hybridization between states with different $S^z$, such as the single-magnon and SIBS. A generalized spin-wave Hamiltonian describes the model's low-energy spectrum as a dilute gas of interacting single-magnon and SIBS quasiparticles, which are treated on equal footing~\cite{Muniz14,bai_hybridized_2020}. To study this problem beyond linear spin-wave theory, we enforce the dilute limit by eliminating states with more than two quasiparticles from the Hilbert space (see SI) and perform an exact diagonalization (ED) of the restricted spin-wave Hamiltonian on a finite lattice of $5\times 5 \times 5$ unit cells (500 spins). As the Hilbert space dimension becomes prohibitively large for ED if we include states with three quasiparticles, our calculation can only account for 1-, 2- and 4-magnon excitations and ignores the 6-magnon states.

The results of our theoretical calculations are shown through the angular average of the frequency-weighted susceptibility $\omega\left[\chi''_{xx}(q=0, \omega) + \chi''_{yy}(q=0, \omega)\right]$ (Fig.~\ref{Fig.TR_TL}{\bf c}) and as the difference between the absorption in the right and the left channels via $\omega\left[\chi''_{+-}(q=0, \omega) - \chi''_{-+}(q=0, \omega)\right]$ (Fig.~\ref{Fig.Ellipticity}{\bf c}). Both plots exhibit a remarkable correspondence with the experimental data (excepting the absence of 6-magnon excitations in the ED). Compared to the linear generalized spin-wave approach of Ref.~\cite{bai_hybridized_2020}, ED includes non-linear interactions that slightly renormalize quasiparticle energies. Therefore, precisely reproducing the energies of the low-energy modes in zero-field requires us to slightly adjust the parameters in Eq.~\ref{eq:hamiltnian}  (see SI) compared to Ref.~\cite{bai_hybridized_2020}. 

\begin{figure*}[t]
\includegraphics[width=2.0\columnwidth]{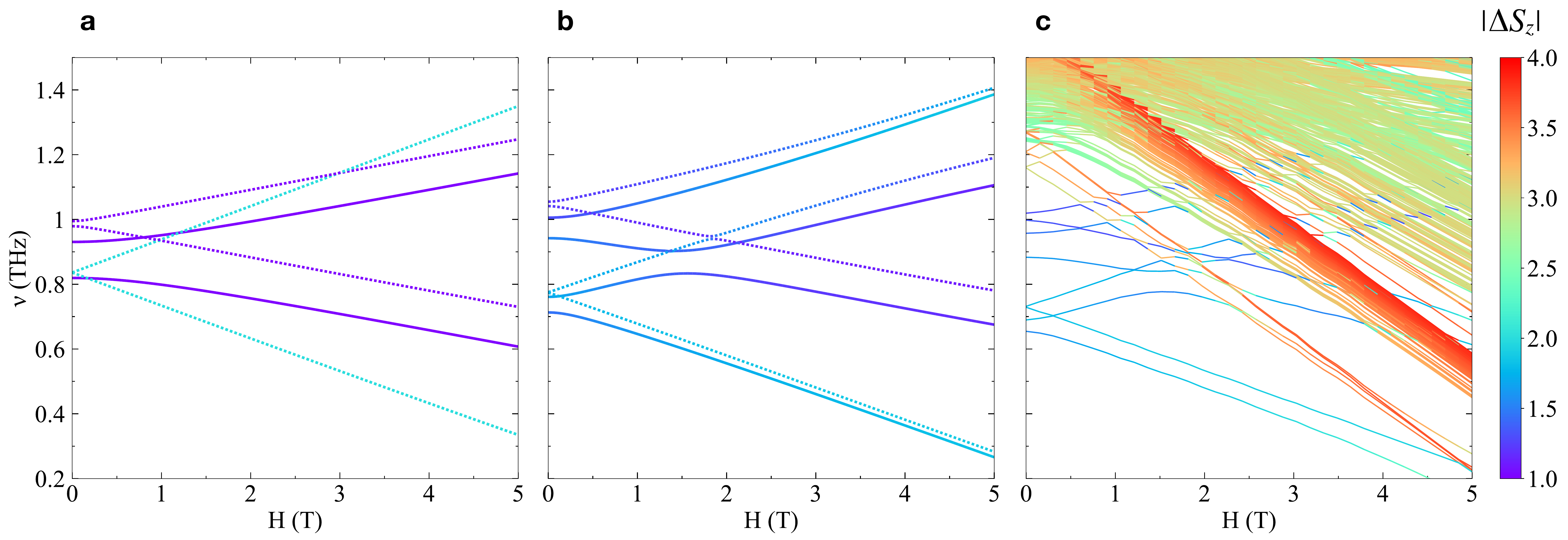}
\caption{\textbf{Theoretical modeling of multi-magnon states in FeI$_2$.}
The value of $\vert\Delta S^z\vert$ calculated using ED of a generalized spin-wave theory Hamiltonian including: \textbf{a.}~only single and two-magnon excitations, without hybridization terms ($J_1^{z \pm} = J_1^{\pm \pm}=0$). Full (dotted) lines correspond to optically active (inactive) branches due to sublattice effects; \textbf{b.}~same as the previous panel but with hybridization effects included; \textbf{c.}~same as previous panel but with the addition of 4-magnon bound states and no distinction between optically active and inactive branches. These plots give an insight on the character of the various magnetic excitations and the hybridization between them.}%
\label{Fig.Theory}%
\end{figure*}

ED calculations give the absolute value of the change in $\boldsymbol{c}$-axis magnetization, $\vert\Delta S^z\vert$, for each excitation branch as a function of frequency and magnetic field (Fig.~\ref{Fig.Theory}). Given that the striped magnetic structure of FeI$_2$ has several sub-lattices, we expect various branches to be present for each magnon sector, some of which optically inactive due to sub-lattice effects. The color-coded expectation value of $\vert\Delta S^z\vert$ shows that in the absence of hybridization ($J_1^{z \pm} = J_1^{\pm \pm}=0$), excitations have a well defined $S^z$ character, with single and two-magnon modes in purple and blue, respectively (Fig.~\ref{Fig.Theory}{\bf a}). The four two-magnon branches are degenerate at zero field as SIBS are completely localized on each lattice site. The Zeeman term splits the zero-field SIBS quartet into two doublets. In the presence of the hybridization terms $J_1^{z \pm}$ and  $J_1^{\pm \pm}$, the two SIBS doublets are split by the $J^{z \pm}$ term via hybridization with \emph{non-degenerate} single-magnon modes (Fig.~\ref{Fig.Theory}{\bf b}). Including 4-magnon excitations yields a rich excitation spectrum with a sequence of hybridization effects between all types of excitations between 1~T and 4~T (Fig.~\ref{Fig.Theory}{\bf c}).

These results highlight that the character of magnetic excitations in FeI$_2$ in an applied magnetic field is profoundly affected by their strong mutual hybridization; most branches have a mixed and changing $\vert\Delta S^z\vert$ character. Therefore, even if the effective \textit{g}-factors give some insight into the nature and degeneracies of the observed resonances, labeling excitations based on their $S^z$ is not possible. Given that excitation branches \#5, \#6 and \#7 from Fig.~\ref{Fig.Ellipticity}{\bf a} have primarily 4-magnon character by comparison to theoretical predictions (Fig.~\ref{Fig.Theory}{\bf c}), the even larger slope of excitation branches \#8 and \#9 indicate an even higher-order character, \textit{{\it i.e.}} these are primarily 6-magnon bound states.
The hybridization of 4- and 6-magnon bound states with lower-order excitations has three important consequences. First, they become detectable in TDTS in the proximity of energy crossings. Second, this explains why the effective \textit{g}-factors of the steeper branches do not reach their hypothetical maximum values, as mentioned above. Finally, it explains why the four low-field branches, labeled \#1 to \#4 in Fig.~\ref{Fig.Ellipticity}{\bf a}, have comparable absorption intensities and similar slopes. These are roughly equal mixtures of single and two-magnon excitations.

In summary, we have used high frequency-resolution time-domain terahertz spectroscopy in a magnetic field to study the spin-anisotropic frustrated compound FeI$_2$ and developed a quantitative understanding of the nature of its low-energy excitations for fields up to around 4~T. Our experiments unraveled a wealth of multi-magnon excitations, including 4-magnon and 6-magnon bound states. ED calculations of a generalized spin-wave Hamiltonian~\cite{bai_hybridized_2020} elucidated spectral contributions from each $\vert\Delta S^z\vert$ sector, enabling a complete understanding of the microscopic character of excitations up to 4-magnons. The strong hybridization that stems from spin-space anisotropy leads to a unique spectroscopic situation where bound states and their interactions can be tracked as a function of the magnetic field. Thus, FeI$_2$ is a promising field-tunable material platform to study fundamental quantum phenomena in magnetism and strongly-interacting few-body models encountered in various contexts such as excitons in semiconductors, cold atoms, and nuclear and particle physics.

\section*{Methods}

\paragraph*{\textbf{Sample preparation}}
Polycrystalline samples of FeI$_2$ were prepared from pure elements according to Ref.~\cite{bai_hybridized_2020}. Samples were ground into fine powders in a glovebox and sealed in thin quartz tubes under vacuum. Large single crystals were subsequently grown by passing the tubes in a graphite crucible with rotating speed 20 RPM/min and pulling rate 10mm/hr through a high-temperature induction Bridgman furnace hosted in the PARADIM facility at Johns Hopkins University. Crystals suitable for time-domain terahertz spectroscopy were cut from larger crystals, resulting in shiny plaquettes of several millimeters long with the \textit{c}-axis perpendicular to the large surface. Powder X-ray diffraction was carried out at room temperature to characterize the crystals~\cite{bai_hybridized_2020}. This compound is highly hygroscopic and degrades within a few seconds when exposed to air, so the sample studied in this work was protected by a layer of photoresist on both sides using a spin coater in a dry glove box, and then mounted on a thin Kapton piece. Another identical Kapton piece was covered on both sides with photoresist to serve as a reference for the THz transmission measurements.

\paragraph*{\textbf{Time-domain terahertz spectroscopy measurements}}
Sample and reference were cooled down to \textit{T}~=~4~K, in the magnetically ordered phase, with magnetic fields up to \textit{H} = 5~T being applied at low temperature. The magnetic field is applied in the Faraday geometry (magnetic field parallel to the propagation of the incident THz beam, see Fig.~\ref{Fig.Experiment}). We use a linearly polarized incident beam and a polarization modulation technique to measure the transmitted THz electric field in both \textit{x} and \textit{y} directions simultaneously~\cite{morris_polarization_2012}. This allows us to determine the transmission coefficients of the sample for left-handed and right-handed circular polarizations (labeled LCP and RCP, the eigenstates of the transmission for this system in such geometry). Indeed, by measuring the complex transmission coefficients \textit{T$_{xx}$} and \textit{T$_{xy}$}, we can determine the coefficients \textit{T$_R$} and \textit{T$_L$} for RCP and LCP light using the change of basis $T_{R,L}$~=~$T_{xx}$~$\pm$~$iT_{xy}$. On this sample, our TDTS experimental set-up enables us to measure accurately in the frequency range of 0.2~THz to 2~THz. A set of measurements of the transmission coefficient in zero field at different temperatures ranging from 3 K up to 300 K confirmed the appearance of magnetic excitations below the N\'{e}el temperature in this frequency range.

To obtain the highest resolution in frequency reachable with our experimental set-up, we need to measure long time scans. Indeed the length of the time trace, labeled $\tau$, directly affects the resolution in frequency after Fourier transform ($\Delta\nu$~=~1/$\tau$). However, such data is affected by the Fabry-Perot effect {\it e.g.} multiple reflections within the crystal which appear in the time trace signal, inducing artificial oscillations in the frequency domain. Therefore, we use the same technique of referencing as in Morris et al.~\cite{morris_hierarchy_2014} to extract the high-resolution spectra. We first reference long time scans at low temperature (4~K) with time scans of the same length at higher temperature (11~K, above the N\'eel temperature) where no sharp feature is observed.

Assuming that the index of refraction is the same at these two temperatures (aside from the absorption due to magnetic excitations), this procedure cancels the Fabry-Perot oscillations. The resulting spectrum is further referenced to a spectrum at \textit{T} = 60K to correct for the broad resonance still present at \textit{T} = 11K, using now short time scans to avoid the Fabry-Perot effect (but giving a high enough resolution to account for the broad resonance at 11K), hence the displayed spectra as follows:
\begin{equation}
T_{R,L} = \frac{T_{R,L}(4K,\nu,\tau_l) }{T_{R,L}(11K,\nu,\tau_l)} \frac{T_{R,L}(11K,\nu,\tau_s) }{T_{R,L}(60K,\nu,\tau_s)},
\end{equation}
with $\tau_{l,s}$ corresponding to long and short time scans respectively. See the Supplemental Information for THz spectra at various temperatures from 4K to 60K. The complex transmission is directly related to the magnetic susceptibiliy ($-\ln(T(\omega)) \propto \omega \chi(q=0,\omega)$), but this process makes it difficult to access accurate quantitative values of the susceptibility (hence the renormalization of the transmission mentioned for Fig.~\ref{Fig.TR_TL}). Nevertheless, the gain in frequency resolution enables the characterization of fine features in the magnetic spectra.

One can also extract the ellipticity of the sample as a function of frequency and field, which quantifies the magnetic dichroism of the material. A complex angle $\theta$ given by $\arctan[(T_R-T_L)/i(T_R+T_L)]$  contains all information on the THz polarization rotation: the real part of this angle $\theta$ is the angle of rotation of the major axes of the elliptically polarized transmitted beam (polar rotating angle) and the imaginary part is the ellipticity.

\bibliographystyle{unsrtnat}
\bibliography{library}

\begin{thebibliography}{24}
\providecommand{\natexlab}[1]{#1}
\providecommand{\url}[1]{\texttt{#1}}
\expandafter\ifx\csname urlstyle\endcsname\relax
  \providecommand{\doi}[1]{doi: #1}\else
  \providecommand{\doi}{doi: \begingroup \urlstyle{rm}\Url}\fi

\bibitem[Torrance and Tinkham(1969)]{torrance_excitation_1969}
J.~B. Torrance and M.~Tinkham.
\newblock Excitation of multiple-magnon bound states in
  {CoCl$_2$}{\textperiodcentered}2{H$_2$O}.
\newblock \emph{Phys. Rev.}, 187\penalty0 (2):\penalty0 595--606, November
  1969.
\newblock \doi{10.1103/PhysRev.187.595}.
\newblock URL \url{https://link.aps.org/doi/10.1103/PhysRev.187.595}.

\bibitem[Hoogerbeets et~al.(1984)Hoogerbeets, Duyneveldt, Phaff, Swuste, and
  Jonge]{hoogerbeets_evidence_1984}
R.~Hoogerbeets, A.~J.~van Duyneveldt, A.~C. Phaff, C.~H.~W. Swuste, and W.~J.
  M.~de Jonge.
\newblock Evidence for magnon bound-state excitations in the quantum chain
  system ({C$_6$H$_{11}$N$_3$}){CuCl$_3$}.
\newblock \emph{J. Phys. C: Solid State Phys.}, 17\penalty0 (14):\penalty0
  2595--2608, May 1984.
\newblock URL \url{https://doi.org/10.1088\%2F0022-3719\%2F17\%2F14\%2F016}.

\bibitem[Broholm et~al.(2020)Broholm, Cava, Kivelson, Nocera, Norman, and
  Senthil]{broholm_quantum_2020}
C~Broholm, RJ~Cava, SA~Kivelson, DG~Nocera, MR~Norman, and T~Senthil.
\newblock Quantum spin liquids.
\newblock \emph{Science}, 367:\penalty0 6475, 2020.
\newblock URL \url{https://science.sciencemag.org/content/367/6475/eaay0668}.

\bibitem[Gingras and McClarty(2014)]{gingras_quantum_2014}
M.~J.~P. Gingras and P.~A. McClarty.
\newblock Quantum spin ice: a search for gapless quantum spin liquids in
  pyrochlore magnets.
\newblock \emph{Rep. Prog. Phys.}, 77\penalty0 (5):\penalty0 056501, May 2014.
\newblock URL \url{https://doi.org/10.1088\%2F0034-4885\%2F77\%2F5\%2F056501}.

\bibitem[Bethe(1931)]{bethe1931theorie}
Hans Bethe.
\newblock Zur theorie der metalle.
\newblock \emph{Zeitschrift f{\"u}r Physik}, 71\penalty0 (3-4):\penalty0
  205--226, 1931.
\newblock URL \url{https://doi.org/10.1007/BF01341708}.

\bibitem[Chauhan et~al.(2020)Chauhan, Mahmood, Changlani, Koohpayeh, and
  Armitage]{chauhan_tunable_2020}
Prashant Chauhan, Fahad Mahmood, Hitesh~J. Changlani, S.~M. Koohpayeh, and
  N.~P. Armitage.
\newblock Tunable magnon interactions in a ferromagnetic spin-1 chain.
\newblock \emph{Phys. Rev. Lett.}, 124\penalty0 (3):\penalty0 037203, January
  2020.
\newblock URL \url{https://link.aps.org/doi/10.1103/PhysRevLett.124.037203}.

\bibitem[Dally et~al.(2020)Dally, Heng, Keselman, Bordelon, Stone, Balents, and
  Wilson]{dally_three-magnon_2020}
Rebecca~L. Dally, Alvin J.~R. Heng, Anna Keselman, Mitchell~M. Bordelon,
  Matthew~B. Stone, Leon Balents, and Stephen~D. Wilson.
\newblock Three-magnon bound state in the quasi-one-dimensional antiferromagnet
  {$\alpha$-NaMnO$_2$}.
\newblock \emph{Phys. Rev. Lett.}, 124\penalty0 (19):\penalty0 197203, May
  2020.
\newblock URL \url{https://link.aps.org/doi/10.1103/PhysRevLett.124.197203}.

\bibitem[Nishida et~al.(2013)Nishida, Kato, and Batista]{Nishida_2013}
Y.~Nishida, Y.~Kato, and C.~Batista.
\newblock {Efimov effect in quantum magnets}.
\newblock \emph{Nat. Phys.}, 9:\penalty0 93--97, 2013.
\newblock URL \url{https://doi.org/10.1038/nphys2523}.

\bibitem[Kato et~al.(2020)Kato, Zhang, Nishida, and Batista]{Kato_2020}
Yasuyuki Kato, Shang-Shun Zhang, Yusuke Nishida, and C.~D. Batista.
\newblock Magnetic field induced tunability of spin {Hamiltonians}: Resonances
  and {Efimov} states in {Yb$_2$Ti$_2$O$_7$}.
\newblock \emph{Phys. Rev. Research}, 2:\penalty0 033024, Jul 2020.
\newblock URL \url{https://link.aps.org/doi/10.1103/PhysRevResearch.2.033024}.

\bibitem[Wang et~al.(2018)Wang, Wu, Yang, Bera, Kamenskyi, Islam, Xu, Law,
  Lake, Wu, et~al.]{wang2018experimental}
Zhe Wang, Jianda Wu, Wang Yang, Anup~Kumar Bera, Dmytro Kamenskyi, ATM~Nazmul
  Islam, Shenglong Xu, Joseph~Matthew Law, Bella Lake, Congjun Wu, et~al.
\newblock Experimental observation of {Bethe} strings.
\newblock \emph{Nature}, 554\penalty0 (7691):\penalty0 219--223, 2018.
\newblock URL \url{https://www.nature.com/articles/nature25466}.

\bibitem[Keselman et~al.(2020)Keselman, Balents, and
  Starykh]{keselman2020dynamical}
Anna Keselman, Leon Balents, and Oleg~A. Starykh.
\newblock Dynamical signatures of quasiparticle interactions in quantum spin
  chains.
\newblock \emph{Phys. Rev. Lett.}, 125:\penalty0 187201, Oct 2020.
\newblock URL \url{https://link.aps.org/doi/10.1103/PhysRevLett.125.187201}.

\bibitem[Yoshida et~al.(2017)Yoshida, Nawa, Ishikawa, Takigawa, Jeong,
  Kr{\"a}mer, Horvati{\'c}, Berthier, Matsui, Goto, et~al.]{yoshida2017spin}
M~Yoshida, K~Nawa, H~Ishikawa, M~Takigawa, Minki Jeong, S~Kr{\"a}mer,
  M~Horvati{\'c}, C~Berthier, K~Matsui, T~Goto, et~al.
\newblock Spin dynamics in the high-field phases of volborthite.
\newblock \emph{Physical Review B}, 96\penalty0 (18):\penalty0 180413, 2017.
\newblock URL
  \url{https://journals.aps.org/prb/abstract/10.1103/PhysRevB.96.180413}.

\bibitem[Pradhan et~al.(2020)Pradhan, Patel, and Trivedi]{pradhan2020two}
Subhasree Pradhan, Niravkumar~D Patel, and Nandini Trivedi.
\newblock Two-magnon bound states in the {Kitaev} model in a [111] field.
\newblock \emph{Physical Review B}, 101\penalty0 (18):\penalty0 180401, 2020.
\newblock URL
  \url{https://journals.aps.org/prb/abstract/10.1103/PhysRevB.101.180401}.

\bibitem[Ward et~al.(2017)Ward, Mena, Bouillot, Kollath, Giamarchi, Schmidt,
  Normand, Kr{\"a}mer, Biner, Bewley, et~al.]{ward2017bound}
Simon Ward, M~Mena, Pierre Bouillot, Corinna Kollath, Thierry Giamarchi,
  KP~Schmidt, B~Normand, KW~Kr{\"a}mer, D~Biner, R~Bewley, et~al.
\newblock Bound states and field-polarized {Haldane} modes in a quantum spin
  ladder.
\newblock \emph{Physical Review Letters}, 118\penalty0 (17):\penalty0 177202,
  2017.
\newblock URL
  \url{https://journals.aps.org/prl/abstract/10.1103/PhysRevLett.118.177202}.

\bibitem[Wulferding et~al.(2020)Wulferding, Choi, Do, Lee, Lemmens, Faugeras,
  Gallais, and Choi]{wulferding_magnon_2020}
Dirk Wulferding, Youngsu Choi, Seung-Hwan Do, Chan~Hyeon Lee, Peter Lemmens,
  Clément Faugeras, Yann Gallais, and Kwang-Yong Choi.
\newblock Magnon bound states versus anyonic {Majorana} excitations in the
  {Kitaev} honeycomb magnet $\alpha$-{RuCl$_3$}.
\newblock \emph{Nature Communications}, 11\penalty0 (1):\penalty0 1603, March
  2020.
\newblock URL \url{https://www.nature.com/articles/s41467-020-15370-1}.

\bibitem[Subrahmanyam(2004)]{subrahmanyam_entanglement_2004}
V.~Subrahmanyam.
\newblock Entanglement dynamics and quantum-state transport in spin chains.
\newblock \emph{Phys. Rev. A}, 69\penalty0 (3):\penalty0 034304, March 2004.
\newblock URL \url{https://link.aps.org/doi/10.1103/PhysRevA.69.034304}.

\bibitem[Fert et~al.(1973)Fert, Gelard, and Carrara]{fert_phase_1973}
A.~R. Fert, J.~Gelard, and P.~Carrara.
\newblock Phase transitions of {FeI$_2$} in high magnetic field parallel to the
  spin direction, static field up to 150 {kOe}, pulsed field up to 250 {kOe}.
\newblock \emph{Solid State Communications}, 13\penalty0 (8):\penalty0
  1219--1223, October 1973.
\newblock URL
  \url{http://www.sciencedirect.com/science/article/pii/0038109873905681}.

\bibitem[Bai et~al.(2020)Bai, Zhang, Dun, Zhang, Huang, Zhou, Stone,
  Kolesnikov, Ye, Batista, and Mourigal]{bai_hybridized_2020}
Xiaojian Bai, Shang-Shun Zhang, Zhiling Dun, Hao Zhang, Qing Huang, Haidong
  Zhou, Matthew~B. Stone, Alexander~I. Kolesnikov, Feng Ye, Cristian~D.
  Batista, and Martin Mourigal.
\newblock Hybridized quadrupolar excitations in the spin-anisotropic frustrated
  magnet {FeI$_2$}.
\newblock \emph{arXiv:2004.05623}, April 2020.
\newblock URL \url{http://arxiv.org/abs/2004.05623}.

\bibitem[Morris et~al.(2012)Morris, Aguilar, Stier, and
  Armitage]{morris_polarization_2012}
C.~M. Morris, R.~Vald{\'e}s Aguilar, A.~V. Stier, and N.~P. Armitage.
\newblock Polarization modulation time-domain terahertz polarimetry.
\newblock \emph{Opt. Express}, 20\penalty0 (11):\penalty0 12303--12317, May
  2012.
\newblock URL
  \url{https://www.osapublishing.org/oe/abstract.cfm?uri=oe-20-11-12303}.

\bibitem[Wiedenmann et~al.(1988)Wiedenmann, Regnault, Burlet, Rossat-Mignod,
  Kound{\'e}, and Billerey]{wiedenmann_neutron_1988}
A.~Wiedenmann, L.~P. Regnault, P.~Burlet, J.~Rossat-Mignod, O.~Kound{\'e}, and
  D.~Billerey.
\newblock A neutron scattering investigation of the magnetic phase diagram of
  {FeI$_2$}.
\newblock \emph{Journal of Magnetism and Magnetic Materials}, 74\penalty0
  (1):\penalty0 7--21, August 1988.
\newblock URL
  \url{http://www.sciencedirect.com/science/article/pii/0304885388901436}.

\bibitem[Morris et~al.(2014)Morris, Vald{\'e}s~Aguilar, Ghosh, Koohpayeh,
  Krizan, Cava, Tchernyshyov, McQueen, and Armitage]{morris_hierarchy_2014}
C.~M. Morris, R.~Vald{\'e}s~Aguilar, A.~Ghosh, S.~M. Koohpayeh, J.~Krizan,
  R.~J. Cava, O.~Tchernyshyov, T.~M. McQueen, and N.~P. Armitage.
\newblock Hierarchy of bound states in the one-dimensional ferromagnetic
  {Ising} chain {CoNb$_2$O$_6$} investigated by high-resolution time-domain
  terahertz spectroscopy.
\newblock \emph{Phys. Rev. Lett.}, 112\penalty0 (13):\penalty0 137403, April
  2014.
\newblock URL \url{https://link.aps.org/doi/10.1103/PhysRevLett.112.137403}.

\bibitem[Petitgrand et~al.(1980)Petitgrand, Brun, and
  Meyer]{petitgrand_magnetic_1980}
D.~Petitgrand, A.~Brun, and P.~Meyer.
\newblock Magnetic field dependence of spin waves and two magnon bound states
  in {FeI$_2$}.
\newblock \emph{Journal of Magnetism and Magnetic Materials}, 15-18:\penalty0
  381--382, January 1980.
\newblock URL
  \url{http://www.sciencedirect.com/science/article/pii/0304885380910975}.

\bibitem[Katsumata et~al.(2000)Katsumata, Yamaguchi, Hagiwara, Tokunaga,
  Mikeska, Goy, and Gross]{katsumata_single-ion_2000}
K.~Katsumata, H.~Yamaguchi, M.~Hagiwara, M.~Tokunaga, H.-J. Mikeska, P.~Goy,
  and M.~Gross.
\newblock Single-ion magnon bound states in an antiferromagnet with strong
  uniaxial anisotropy.
\newblock \emph{Phys. Rev. B}, 61\penalty0 (17):\penalty0 11632--11636, May
  2000.
\newblock URL \url{https://link.aps.org/doi/10.1103/PhysRevB.61.11632}.

\bibitem[Muniz et~al.(2014)Muniz, Kato, and Batista]{Muniz14}
Rodrigo~A Muniz, Yasuyuki Kato, and Cristian~D Batista.
\newblock Generalized spin-wave theory: Application to the
  bilinear--biquadratic model.
\newblock \emph{Progress of Theoretical and Experimental Physics},
  2014:\penalty0 8, 2014.
\newblock URL \url{https://doi.org/10.1093/ptep/ptu109}.

\end{thebibliography}


\begin{thebibliography}{6}
\providecommand{\natexlab}[1]{#1}
\providecommand{\url}[1]{\texttt{#1}}
\expandafter\ifx\csname urlstyle\endcsname\relax
  \providecommand{\doi}[1]{doi: #1}\else
  \providecommand{\doi}{doi: \begingroup \urlstyle{rm}\Url}\fi

\bibitem[Bai et~al.(2020)Bai, Zhang, Dun, Zhang, Huang, Zhou, Stone,
  Kolesnikov, Ye, Batista, and Mourigal]{bai_hybridized_2020}
Xiaojian Bai, Shang-Shun Zhang, Zhiling Dun, Hao Zhang, Qing Huang, Haidong
  Zhou, Matthew~B. Stone, Alexander~I. Kolesnikov, Feng Ye, Cristian~D.
  Batista, and Martin Mourigal.
\newblock Hybridized quadrupolar excitations in the spin-anisotropic frustrated
  magnet {FeI$_2$}.
\newblock \emph{arXiv:2004.05623}, April 2020.
\newblock URL \url{http://arxiv.org/abs/2004.05623}.

\bibitem[Gagliano and Balseiro(1987)]{Gagliano87}
E.~R. Gagliano and C.~A. Balseiro.
\newblock Dynamical properties of quantum many-body systems at zero
  temperature.
\newblock \emph{Phys. Rev. Lett.}, 59:\penalty0 2999--3002, Dec 1987.
\newblock URL \url{https://link.aps.org/doi/10.1103/PhysRevLett.59.2999}.

\bibitem[Lanczos(1950)]{lanczos1950iteration}
Cornelius Lanczos.
\newblock \emph{An iteration method for the solution of the eigenvalue problem
  of linear differential and integral operators}.
\newblock United States Governm. Press Office Los Angeles, CA, 1950.

\bibitem[Petitgrand et~al.(1980)Petitgrand, Brun, and
  Meyer]{petitgrand_magnetic_1980}
D.~Petitgrand, A.~Brun, and P.~Meyer.
\newblock Magnetic field dependence of spin waves and two magnon bound states
  in {FeI$_2$}.
\newblock \emph{Journal of Magnetism and Magnetic Materials}, 15-18:\penalty0
  381--382, January 1980.
\newblock URL
  \url{http://www.sciencedirect.com/science/article/pii/0304885380910975}.

\bibitem[Katsumata et~al.(2000)Katsumata, Yamaguchi, Hagiwara, Tokunaga,
  Mikeska, Goy, and Gross]{katsumata_single-ion_2000}
K.~Katsumata, H.~Yamaguchi, M.~Hagiwara, M.~Tokunaga, H.-J. Mikeska, P.~Goy,
  and M.~Gross.
\newblock Single-ion magnon bound states in an antiferromagnet with strong
  uniaxial anisotropy.
\newblock \emph{Phys. Rev. B}, 61\penalty0 (17):\penalty0 11632--11636, May
  2000.
\newblock URL \url{https://link.aps.org/doi/10.1103/PhysRevB.61.11632}.

\bibitem[Pan et~al.(2014)Pan, Kim, Ghosh, Morris, Ross, Kermarrec, Gaulin,
  Koohpayeh, Tchernyshyov, and Armitage]{pan2014low}
LiDong Pan, Se~Kwon Kim, A~Ghosh, Christopher~M Morris, Kate~A Ross, Edwin
  Kermarrec, Bruce~D Gaulin, SM~Koohpayeh, Oleg Tchernyshyov, and NP~Armitage.
\newblock Low-energy electrodynamics of novel spin excitations in the quantum
  spin ice {Yb$_2$Ti$_2$O$_7$}.
\newblock \emph{Nature communications}, 5\penalty0 (1):\penalty0 1--7, 2014.
\newblock URL \url{https://www.nature.com/articles/ncomms5970}.

\end{thebibliography}

\section{Data Availability}
Data in this manuscript is available from the corresponding author upon reasonable request.

\section{Acknowledgments}
Work at Johns Hopkins University (A.L, N.P.A.) was supported as part of the Institute for Quantum Matter, an EFRC funded by the DOE BES under DE-SC0019331. The work at Georgia Tech (X.B., Z.L.D., M.M.) was supported by the U.S. Department of Energy, Office of Science, Basic Energy Sciences, Materials Sciences and Engineering Division under award DE-SC-0018660. Growth of FeI$_2$ crystals is based upon work supported by the National Science Foundation (Platform for the Accelerated Realization, Analysis, and Discovery of Interface Materials (PARADIM)) under Cooperative Agreement No. DMR-1539918. Work by H.Z. was supported by the U.S. Department of Energy (DOE), Office of Science, Basic Energy Sciences, Materials Sciences and Engineering Division. The work at the University of Tennessee (C.D.B. and S-S.Z.) was partially funded by the U.S. Department of Energy, Office of Basic Energy Sciences.

\section{Author Contributions}

A.L. performed the experiments and analyzed the THz data.  S.S.Z, H.Z., and C.B. performed the theoretical calculations. X.B., Z.L.D., and W.A.P grew the crystals.  C.D.B, M.M., and N.P.A supervised the project. A.L., S.S.Z., X.B., H.Z., C.B., M.M. and N.P.A. contributed to the discussion and to the write-up of the manuscript.



\end{document}


\title{Supplemental Information for ``Observation of 4- and 6-magnon bound-states in the spin-anisotropic frustrated antiferromagnet FeI$_2$"}

\author{Ana\"elle Legros}
	\thanks{These authors contributed equally to this work}
	\affiliation{The Institute for Quantum Matter and the Department of Physics and Astronomy, The Johns Hopkins University, Baltimore, MD 21218, USA}
\author{Shang-Shun Zhang}
	\thanks{These authors contributed equally to this work}
	\affiliation{Department of Physics and Astronomy, University of Tennessee, Knoxville, TN 37996, USA}
\author{Xiaojian Bai}
	\affiliation{School of Physics, Georgia Institute of Technology, Atlanta, GA 30332, USA}
	\affiliation{Neutron Scattering Division, Oak Ridge National Laboratory, Oak Ridge, TN 37831, USA}
\author{Hao Zhang}
	\affiliation{Department of Physics and Astronomy, University of Tennessee, Knoxville, TN 37996, USA}
	\affiliation{Materials Science and Technology Division, Oak Ridge National Laboratory, Oak Ridge, TN 37831, USA}
\author{Zhiling Dun}
	\affiliation{School of Physics, Georgia Institute of Technology, Atlanta, GA 30332, USA}
\author{W. Adam Phelan}
	\affiliation{PARADIM, Department of Chemistry, The Johns Hopkins University, Baltimore, MD 21218, USA}
\author{Cristian D. Batista}
	\affiliation{Department of Physics and Astronomy, University of Tennessee, Knoxville, TN 37996, USA}
\author{Martin Mourigal}
	\affiliation{School of Physics, Georgia Institute of Technology, Atlanta, GA 30332, USA}
\author{N. P. Armitage}
	\email[To whom correspondence should be addressed, ]{npa@jhu.edu}
	\affiliation{The Institute for Quantum Matter and the Department of Physics and Astronomy, The Johns Hopkins University, Baltimore, MD 21218, USA}

\maketitle

\tableofcontents

\section{Generalized spin-wave theory}

The FeI$_{2}$ compound under external field is described by the effective
$S=1$ spin model, 
\begin{eqnarray}
{\cal H}=\sum_{\langle ij\rangle}\sum_{\mu\nu}{S}_{i}^{\mu}{\cal J}_{ij}^{\mu\nu}{S}_{j}^{\nu}-D\sum_{i}Q_{i}^{zz}-\mu_{0}\mu_{B}gH\sum_{i}S_{i}^{z},\label{eq:hamiltnian}
\end{eqnarray}
where ${S}_{i}^{\mu},\mu=x,y,z$ is the spin-$1$ operator and the
single-ion anisotropy term is proportional to the $(zz)$ component of quadrupolar moment $Q_{i}^{\mu\nu}=({S}_{i}^{\mu}{S}_{i}^{\nu}+{S}_{i}^{\nu}{S}_{i}^{\mu})/2-2/3{\delta^{\mu\nu}}$
(symmetric traceless components of ${\bm{S}}_{i}\otimes{\bm{S}}_{i}$).
The spin-exchange tensor ${\cal J}_{ij}^{\mu\nu}$ is described in
the main text and a detailed symmetry analysis is provided in Ref.~\onlinecite{bai_hybridized_2020}.
A distinctive property of FeI$_{2}$ is that its low-energy modes
include both dipolar and quadrupolar fluctuations because of a subtle
balance between the magnitude of the exchange interaction and the
single-ion anisotropy. Consequently, the usual SU(2) spin-wave theory
must be generalized to include both types of low-energy modes. The
three components of the magnetization and the five components of the
quadrupolar moment generate the {SU(3) unitary} transformations
in the 3-dimensional Hilbert space of a $S=1$ spin. Correspondingly,
an SU(3) spin-wave theory can simultaneously account for the low-energy
dipolar and quadrupolar fluctuations of FeI$_{2}$. This generalization
can be implemented by introducing the SU(3) Schwinger boson representation
of the spin operators ${S}_{i}^{\mu}={\psi}_{i}^{\dagger}L^{\mu}{\psi}_{i}$~\cite{Muniz_2014},
where ${\psi}_{i}=(b_{i,+1},b_{i,0},b_{i,-1})$ and 
\begin{eqnarray}
L^{x}=\left(\begin{array}{ccc}
0 & -\frac{i}{\sqrt{2}} & 0\\
\frac{1}{\sqrt{2}} & 0 & \frac{1}{\sqrt{2}}\\
0 & \frac{1}{\sqrt{2}} & 0
\end{array}\right),L^{y}=\left(\begin{array}{ccc}
0 & \frac{1}{\sqrt{2}} & 0\\
\frac{i}{\sqrt{2}} & 0 & -\frac{i}{\sqrt{2}}\\
0 & \frac{i}{\sqrt{2}} & 0
\end{array}\right),L^{z}=\left(\begin{array}{ccc}
1 & 0 & 0\\
0 & 0 & 0\\
0 & 0 & -1
\end{array}\right).
\end{eqnarray}
The number of Schwinger bosons per site, $\sum_{m}b_{im}^{\dagger}b_{im}=2S=1$
is determined by the spin size $S=1$.

To compute the excitation spectrum, it is convenient to work in the
local reference frame defined by the SU(3) rotation 
\begin{equation}
\left(\begin{array}{c}
\beta_{i,+1}\\
\beta_{i,0}\\
\beta_{i,-1}
\end{array}\right)=U_{i}^{\dagger}\cdot\left(\begin{array}{c}
b_{i\uparrow}\\
b_{i0}\\
b_{i\downarrow}
\end{array}\right),\label{eq:su3_rotation-1}
\end{equation}
where $U_{i}\equiv A_{1}\cdot A_{2}$ with 
\begin{eqnarray}
A_{1} & \equiv & \left(\begin{array}{ccc}
\sin(\theta)\cos(\phi)e^{i\alpha_{1}} & \cos(\theta)\cos(\phi)e^{i\alpha_{1}} & -\sin(\phi)e^{-i(\alpha_{2}+\alpha_{3})}\\
\sin(\theta)\sin(\phi)e^{i\alpha_{2}} & \cos(\theta)\sin(\phi)e^{i\alpha_{2}} & \cos(\phi)e^{-i(\alpha_{1}+\alpha_{3})}\\
\cos(\theta)e^{i\alpha_{3}} & -\sin(\theta)e^{i\alpha_{3}} & 0
\end{array}\right),
\end{eqnarray}
and 
\begin{eqnarray}
A_{2} & \equiv & \left(\begin{array}{ccc}
1 & 0 & 0\\
0 & \cos(\chi)e^{-i\beta_{1}} & \sin(\chi)e^{i(\beta_{2}-\alpha_{1}-\alpha_{2}-\alpha_{3})}\\
0 & -\sin(\chi)e^{-i(\beta_{2}-\alpha_{1}-\alpha_{2}-\alpha_{3})} & \cos(\chi)e^{i\beta_{1}}
\end{array}\right).
\end{eqnarray}
Up to a phase factor, the first column of $U$ is the optimal mean
field state $\rvert\psi_{i}\rangle=\beta_{i,+1}^{\dagger}\rvert\emptyset\rangle$.  The other two flavors ($m=0,-1$) of $\beta_{i,m}$ are defined by
the second and third columns of $U_{i}$, which are orthogonal to
the first column. For $U_{i}\in SU(2)$, the boson $\beta_{i,m}$
carries a quantized angular momentum $m=1$ along the quantization
axis $\hat{{\bm{n}}}_{i}=\langle\psi_{c}\rvert{\bm{S}}_{i}\rvert\psi_{c}\rangle$.  The condensation of the $\beta_{i,+1}$ boson gives rise to the local
dipole moments along the direction $\hat{{\bm{n}}}_{i}$. The $\beta_{i,-1}$
boson changes the angular momentum by $-2$ relative to the local
quantization axis, \textit{i.e.}, it generates a local quadrupolar
fluctuation. Unless there is a finite hybridization between the $\beta_{i,0}$
and $\beta_{i,-1}$ bosons, quadrupolar excitations remain invisible
to terahertz (THz) spectroscopy and inelastic neutron scattering.
In fact, although the rotation matrix $U_{i}$ can be approximated
by an SU(2) matrix in FeI$_{2}$, the normal modes are strong hybridizations
of the $\beta_{i,0}$ and $\beta_{i,-1}$ bosons, enabling the observation
of predominantly quadrupolar excitations with THz spectroscopy and
inelastic neutron scattering.

The observed magnetic order in FeI$_{2}$ is described by a condensation
of bosons in the single-particle state 
\begin{eqnarray}
\rvert\psi_{i}\rangle & = & \left(\begin{array}{c}
e^{i\alpha_{1}(i)}\cos[\phi(i)]\sin[\theta(i)]\\
e^{i\alpha_{2}(i)}\sin[\phi(i)]\sin[\theta(i)]\\
\cos[\theta(i)]
\end{array}\right)
\end{eqnarray}
on the basis of $\{\rvert i,+1\rangle,\rvert i,0\rangle,\rvert i,-1\rangle\}$,
where $\rvert i,m\rangle=b_{i,m}^{\dagger}\rvert\emptyset\rangle$
with $\rvert\emptyset\rangle$ being the vacuum of Schwinger bosons.
The parameters $\theta(i),\phi(i),\alpha_{1}(i),\alpha_{2}(i)$ are
determined by minimizing the mean field energy $\prod_{i=1}^{N}\langle\psi_{i}\rvert{\cal H}\prod_{i=1}^{N}\rvert\psi_{i}\rangle$
with $N$ being the number of lattice sites. The optimal mean field
state $\rvert\psi_{i}\rangle$ is an SU(3) coherent state, which in
general cannot be obtained from the \textit{fully polarized} state
$(1,0,0)^{T}$ by applying an SU(2) rotation. For an SU(2) spin coherent
state, the length of dipole moment is $d=\sqrt{\sum_{\mu}\langle{S}_{i}^{\mu}\rangle^{2}}=1$.
However, the optimal mean field state for FeI$_{2}$ has $d<1$, implying
that it is not an SU(2) spin coherent state. The deviation arises
from the $J_{1}^{z\pm}$ term that breaks the axial symmetry about
the $c$-axis. The net result is a $10^\circ$ canting of the magnetic
moments away from the $c$-axis and a small reduction in the magnitude
of the moment. Nevertheless, the magnitude of the moment is very close
to one, $d\simeq1$, as anticipated from the rather strong easy-axis
single-ion anisotropy.

Since the $\beta_{i,+1}$ boson is macroscopically occupied, $\langle\beta_{i,+1}\rangle=\langle\beta_{i,+1}^{\dagger}\rangle\simeq\sqrt{M}$
($M=1$ for the case under consideration), we assume that $\langle\beta_{i,0}^{\dagger}\beta_{i,0}\rangle,\langle\beta_{i,-1}^{\dagger}\beta_{i,-1}\rangle\ll M$.
This assumption justifies an expansion in the small parameter $1/M$:
\begin{eqnarray}
\beta_{i,+1} & = & \beta_{i,+1}^{\dagger}=\sqrt{M-\beta_{i,0}^{\dagger}\beta_{i,0}-\beta_{i,-1}^{\dagger}\beta_{i,-1}}\nonumber \\
 & \simeq & \sqrt{M}\left[1-\frac{1}{2M}\beta_{i,0}^{\dagger}\beta_{i,0}-\frac{1}{2M}\beta_{i,-1}^{\dagger}\beta_{i,-1}+{\cal O}\left(\frac{1}{M^{2}}\right)\right],
\end{eqnarray}
that leads to the following expansion of the dipolar and quadrupolar
operators 
\begin{eqnarray}
{S}_{i}^{\mu} & = & M{\cal S}_{c}^{\mu}(i)+\sqrt{M}\sum_{m\neq1}\left({\cal S}_{1m}^{\mu}(i)\beta_{i,m}+h.c.\right)+\sum_{m,n\neq1}{\cal S}_{mn}^{\mu}(i)\beta_{i,m}^{\dagger}\beta_{i,n}+{\cal O}\left(\frac{1}{\sqrt{M}}\right),\label{eq:spin}
\end{eqnarray}
\begin{eqnarray}
{Q}_{i}^{zz} & = & M{\cal Q}_{c}^{zz}(i)+\sqrt{M}\sum_{m\neq1}\left({\cal Q}_{1m}^{zz}(i)\beta_{i,m}+h.c.\right)+\sum_{m,n\neq1}{\cal Q}_{mn}^{zz}(i)\beta_{i,m}^{\dagger}\beta_{i,n}+{\cal O}\left(\frac{1}{\sqrt{M}}\right),\label{eq:quadrupole}
\end{eqnarray}
where 
\begin{eqnarray}
{\cal S}_{c}^{\mu}(i) & = & \left(U_{i}^{\dagger}L^{\mu}U_{i}\right)_{11},{\cal Q}_{c}^{zz}(i)=\left(U_{i}^{\dagger}O^{zz}U_{i}\right)_{11},\label{eq:ob1}\\
{\cal S}_{1m}^{\mu}(i) & = & \left(U_{i}^{\dagger}L^{\mu}U_{i}\right)_{1m},{\cal Q}_{1m}^{zz}(i)=\left(U_{i}^{\dagger}O^{zz}U_{i}\right)_{1m},\label{eq:ob2}\\
{\cal S}_{mn}^{\mu}(i) & = & \left(U_{i}^{\dagger}L^{\mu}U_{i}\right)_{mn}-\left(U_{i}^{\dagger}L^{\mu}U_{i}\right)_{11}\delta_{mn},\label{eq:ob3}\\
{\cal Q}_{mn}^{zz}(i) & = & \left(U_{i}^{\dagger}O^{zz}U_{i}\right)_{mn}-\left(U_{i}^{\dagger}O^{zz}U_{i}\right)_{11}\delta_{mn},\label{eq:ob4}
\end{eqnarray}
with {$O^{zz}=(L^{z})^{2}$}. Note that the variables defined in
Eqs. (\ref{eq:ob1}-\ref{eq:ob4}) depend only on the sublattice index
because of the translation symmetry of the magnetic structure. By
applying the above formula, we obtain the $1/M$ expansion of the
spin Hamiltonian 
\begin{eqnarray}
{\cal H} & = & M^{2}{\cal E}^{(0)}+M{\cal H}^{(2)}+M^{1/2}{\cal H}^{(3)}+M^{0}{\cal H}^{(4)}+{\cal O}(M^{-1}),
\end{eqnarray}
where terms $\propto M^{3/2}$ are not present because we are expanding
around the mean-field state that minimizes the classical energy (${\cal E}^{(0)}$).
The other terms represent the generalized linear spin wave Hamiltonian
\begin{align}
{\cal H}^{(2)} & =\sum_{\langle ij\rangle}\sum_{\mu\nu}{\cal J}_{ij}^{\mu\nu}\sum_{m,n\neq1}\Bigg[{\cal S}_{c}^{\mu}(i){\cal S}_{mn}^{\nu}(j)\beta_{j,m}^{\dagger}\beta_{j,n}+{\cal S}_{mn}^{\mu}(i){\cal S}_{c}^{\nu}(j)\beta_{i,m}^{\dagger}\beta_{i,n}+{\cal S}_{m1}^{\mu}(i){\cal S}_{1n}^{\nu}(j)\beta_{i,m}^{\dagger}\beta_{j,n}\nonumber \\
 & +{\cal S}_{1m}^{\mu}(i){\cal S}_{n1}^{\nu}(j)\beta_{i,m}\beta_{j,n}^{\dagger}+{\cal S}_{1m}^{\mu}(i){\cal S}_{1n}^{\nu}(j)\beta_{i,m}\beta_{j,n}+{\cal S}_{m1}^{\mu}(i){\cal S}_{n1}^{\nu}(j)\beta_{i,m}^{\dagger}\beta_{j,n}^{\dagger}\Bigg]\nonumber \\
 & -D\sum_{i}\sum_{m,n\neq1}{\cal Q}_{mn}^{zz}\beta_{i,m}^{\dagger}\beta_{i,n}-\mu_{0}\mu_{B}gH\sum_{i}\sum_{m,n\neq1}{\cal S}_{mn}^{z}(i)\beta_{i,m}^{\dagger}\beta_{i,n},
\end{align}
the cubic interaction terms, 
\begin{align}
{\cal H}^{(3)} & =\sum_{\langle ij\rangle}\Bigg[\sum_{m,n\neq1}\left(V_{1}^{m}(i,j)\beta_{j,n}^{\dagger}\beta_{j,n}\beta_{j,m}+V_{1}^{m}(j,i)\beta_{i,n}^{\dagger}\beta_{i,n}\beta_{i,m}+h.c.\right)\nonumber \\
 & +\sum_{l,m,n\neq1}\left(V_{2}^{lmn}(i,j)\beta_{j,m}^{\dagger}\beta_{j,n}\beta_{i,l}+V_{2}^{lmn}(j,i)\beta_{i,m}^{\dagger}\beta_{i,n}\beta_{j,l}+h.c.\right)\Bigg]\nonumber \\
 & +\frac{1}{2M}\sum_{i}\sum_{m,n\neq1}\left(\left(D{\cal Q}_{1m}^{zz}(i)+\mu_{0}\mu_{B}gH{\cal S}_{1m}^{z}(i)\right)\beta_{i,n}^{\dagger}\beta_{i,n}\beta_{i,m}+h.c.\right),
\end{align}
where $V_{1}^{m}(i,j)=-\frac{1}{2}{\cal S}_{c}^{\mu}(i){\cal J}_{ij}^{\mu\nu}{\cal S}_{1m}^{\nu}(j)$,
$V_{2}^{lmn}(i,j)={\cal S}_{1l}^{\mu}(i){\cal J}_{ij}^{\mu\nu}{\cal S}_{mn}^{\nu}(j)$,
and the quartic interaction terms 
\begin{align}
{\cal H}^{(4)} & =-\frac{1}{2}\sum_{\langle ij\rangle}\sum_{m,s,r\neq1}\left(U_{1}^{ms}(i,j)\beta_{j,r}^{\dagger}\beta_{j,r}\beta_{j,s}\beta_{i,m}+U_{2}^{ms}(i,j)\beta_{j,s}^{\dagger}\beta_{j,r}^{\dagger}\beta_{j,r}\beta_{i,m}+h.c.\right)\nonumber \\
 & +\sum_{\langle ij\rangle}\sum_{m,n,s,r\neq1}U_{3}^{mn;sr}(i,j)\beta_{i,m}^{\dagger}\beta_{i,n}\beta_{j,s}^{\dagger}\beta_{j,r}\nonumber \\
 & -\frac{1}{2}\sum_{\langle ij\rangle}\sum_{m,n,s\neq1}\left(U_{1}^{sm}(j,i)\beta_{i,n}^{\dagger}\beta_{i,n}\beta_{i,m}\beta_{j,s}+U_{2}^{sm}(j,i)\beta_{i,m}^{\dagger}\beta_{i,n}^{\dagger}\beta_{i,n}\beta_{j,s}+h.c.\right)
\end{align}
where $U_{1}^{ms}(i,j)=\sum_{\mu\nu}{\cal S}_{1m}^{\mu}(i){\cal J}_{ij}^{\mu\nu}{\cal S}_{1s}^{\nu}(j)\equiv U_{1}^{sm}(j,i)$,
$U_{2}^{ms}(i,j)=\sum_{\mu\nu}{\cal S}_{1m}^{\mu}(i){\cal J}_{ij}^{\mu\nu}{\cal S}_{1s}^{\nu*}(j)\equiv\left(U_{2}^{sm}(j,i)\right)^{*}$
and $U_{3}^{mn;sr}(i,j)=\sum_{\mu\nu}{\cal S}_{mn}^{\mu}(i){\cal J}_{ij}^{\mu\nu}{\cal S}_{sr}^{\nu}(j)\equiv U_{3}^{sr;mn}(j,i)$.

The non-interacting part of the spin wave Hamiltonian ${\cal H}^{(2)}$
can be fully diagonalized in momentum space via a Bogoliubov transformation:
\begin{eqnarray}
{\cal H}^{(2)} & = & \sum_{a}\varepsilon_{a}\gamma_{a}^{\dagger}\gamma_{a},
\end{eqnarray}
where $a\equiv(n_{a},\bm{q}_{a})$ with $n_{a}$ the band index and
$\bm{q}_{a}$ the momentum.  The $\gamma$ operator is defined through
\begin{eqnarray}
\left(\begin{array}{c}
\beta_{(\alpha,\bm{q}),\sigma}\\
\beta_{(\alpha,\bar{\bm{q}}),\sigma}^{\dagger}
\end{array}\right) & = & W_{(\alpha,\sigma),n}(\bm{q})\left(\begin{array}{c}
\gamma_{n,\bm{q}}\\
\gamma_{n,\bar{\bm{q}}}^{\dagger}
\end{array}\right),
\end{eqnarray}
where $\beta_{(\alpha,{\bm{q}})\sigma}=N_{uc}^{-1/2}\sum_{\bm{r}}e^{i\bm{q}\cdot\bm{r}}\beta_{(\alpha,\bm{r})\sigma}$
and $W_{(\alpha,\sigma),n}(\bm{q})$ is a paraunitary matrix. The
non-interacting modes include four magnon states and four single-ion
bound states ($2$-magnon bound states). The interaction between the
$\gamma$-particles is obtained from by applying the above-mentioned
transformation to ${\cal H}^{(3)}$ and ${\cal H}^{(4)}$, which formally
leads to 
\begin{eqnarray}
{\cal H}^{(3)} & = & \frac{1}{2!\sqrt{N_{uc}}}\sum_{abc}\delta(\bm{q}_{a}+\bm{q}_{b}+\bm{q}_{c}-\bm{G})V_{abc}\gamma_{\bar{a}}^{\dagger}\gamma_{b}\gamma_{c}+h.c.,\\
{\cal H}^{(4)} & = & \frac{1}{2!2!N_{uc}}\sum_{abcd}\delta(\bm{q}_{a}+\bm{q}_{b}+\bm{q}_{c}+\bm{q}_{d}-\bm{G})U_{abcd}\gamma_{\bar{a}}^{\dagger}\gamma_{\bar{b}}^{\dagger}\gamma_{c}\gamma_{d},
\end{eqnarray}
where $\bar{a}\equiv(n_{a},-\bm{q}_{a})$, and the vertex functions
have been symmetrized relative to permutations of the particle indexes.
Terms of the form $\propto\gamma^{\dagger}\gamma^{\dagger}\gamma^{\dagger},\gamma^{\dagger}\gamma^{\dagger}\gamma^{\dagger}\gamma$ and their hermitian conjugates are ignored because they do not contribute to the low-energy subspace that we introduce below.


The THz spectra reported here show clear evidence of the formation of $n$-magnon bound states with $n=4,6$. These bound states are formed by the binding of $2$ or $3$ single-ion bound states via the spin exchange interactions.  As mentioned in the main text, the tendency for such bound state formation is high because of the flat dispersion of the $n=2$ single-ion bound states.  This non-perturbative effect of the interaction terms can only be captured by summing ladder diagrams up to infinite order (exact solution of two-body problem).  This can be achieved by diagonalizing ${\cal H}$ restricted to the truncated subspace ${\cal S}_{1,2}$ with number of $\gamma$ quasi-particles $n\leq2$. While this exact diagonalization can only be performed in a finite size cluster, the numerical cost of diagonalizing ${\cal P}_{1,2}{\cal H}{\cal P}_{1,2}$ for a fixed center of mass momentum ${\bm{K}}$ (${\bm{K}}=0$ for THz spectroscopy) increases \emph{linearly} in the system size. Here, ${\cal P}_{1,2}$ is the projector on the low-energy subspace ${\cal S}_{1,2}$. Besides capturing the formation of $4$-magnon bound states, this approach also captures the hybridization of these new low-energy modes with the single $\gamma$-modes.  In other words, the approach cannot capture the formation of $6$-magnon bound states because those states correspond to three-body states of the $\gamma$ quasi-particles that have a strong two-magnon character. The description of the latter mode requires at least three non-interacting particles which is too expensive in the current model. The subspace ${\cal S}_{1,2}$ is spanned by the basis $\{\rvert i\rangle,\rvert i\leq j\rangle\}$ with $\rvert i\rangle=\gamma_{i}^{\dagger}\rvert\emptyset\rangle$ and $\rvert i\leq j\rangle=\zeta_{ij}\gamma_{i}^{\dagger}\gamma_{j}^{\dagger}\rvert\emptyset\rangle$, where $\rvert\emptyset\rangle$ refers to the vacuum of the $\gamma$ quasi-particles and $\zeta_{i\neq j}=1$ and $\zeta_{i=j}=1/\sqrt{2!}$ are normalization factors. The projection of ${\cal H}$ into the subspace ${\cal S}_{1,2}$ takes the matrix form 
\begin{eqnarray}
{\cal P}_{1,2}{\cal H}{\cal P}_{1,2} & = & \left(\begin{array}{cc}
{\cal H}_{11} & {\cal H}_{12}\\
h.c. & {\cal H}_{22}
\end{array}\right),\label{eq:H_1}
\end{eqnarray}
with matrix elements 
\begin{eqnarray*}
{\cal H}_{11}^{i,j}=\delta_{ij}\varepsilon_{i},\;\;\;{\cal H}_{12}^{i,j\leq k}=\frac{1}{\sqrt{N_{uc}}}V_{\bar{i}jk}\zeta_{j,k},\;\;\;{\cal H}_{22}^{i\leq j,k\leq l}=\delta_{ik}\delta_{jl}(\varepsilon_{i}+\varepsilon_{j})\zeta_{ij}^{2}+\frac{1}{N_{uc}}U_{\bar{i}\bar{j}kl}\zeta_{ij}\zeta_{kl}.
\end{eqnarray*}
The diagonalization of ${\cal P}_{1,2}{\cal H}{\cal P}_{1,2}$ reveals
that the spectrum includes four energy levels below the two-particle
continuum with a strong $4$-magnon character, which are identified
as the $4$-magnon bound states.

The THz absorption spectrum is obtained by computing 
\begin{eqnarray}
I(\omega) & \propto & \omega\left(\chi_{xx}^{\prime\prime}({\bm{q}}={\bm{0}},\omega)+\chi_{yy}^{\prime\prime}({\bm{q}}={\bm{0}},\omega)\right),
\end{eqnarray}
where $\chi_{\mu\nu}^{\prime\prime}({\bm{q}}={\bm{0}},\omega)$ is
the imaginary part of the uniform magnetic susceptibility. $\chi_{\mu\nu}^{\prime\prime}({\bm{q}},\omega)$
is related to the dynamical spin structure factor $S_{\mu\nu}({\bm{q}},\omega>0)$
through the fluctuation-dissipation theorem, which takes the form
$2\chi_{\mu\nu}^{\prime\prime}({\bm{q}},\omega>0)=S_{\mu\nu}({\bm{q}},\omega>0)$
at zero temperature, where 
\begin{align}
S_{\mu\nu}({\bm{q}}={\bm{0}},\omega) & =\int_{-\infty}^{\infty}dte^{i\omega t}\frac{1}{N}\sum_{ij}\langle\emptyset\ \rvert S_{\bm{r}_{j}}^{\mu}(t)S_{{\bf r}_{i}}^{\nu}(0)\rvert\emptyset\ \rangle\nonumber \\
 & =\int_{-\infty}^{\infty}dte^{i\omega t}\frac{1}{4}\sum_{\alpha\beta}\langle\emptyset\ \rvert S_{\alpha,{\bf 0}}^{\mu}(t)S_{\beta,\bm{0}}^{\nu}(0)\rvert\emptyset\rangle,
\end{align}
$N=4N_{uc}$ is the total number of lattice sites, and $S_{\alpha,{\bf 0}}^{\mu}(t)=(1/\sqrt{N_{uc}})\sum_{\bm{r}\in\alpha}S_{\bm{r}}^{\mu}(t)$ is the uniform component of the spin operators on the sublattice $\alpha$. This correlation function has been computed on finite lattice of 
$5 \times 5 \times 5$ unit cells (500 spins) by using the 
continued-fraction method~\cite{Gagliano87}  based on the Lanczos algorithm~\cite{lanczos1950iteration}. This size is large enough to capture the 4-magnon bound states because their linear size is of the order of one lattice space owing to the very large effective mass of the two-magnon bound sates.

\begin{table}[h]
\centering
\begin{tabular}{c|c|c|c|c|c}
\hline
$J^{\pm}_1$ (meV)  & $J^{zz}_1$   & $J^{\pm\pm}_1$   & $J^{z\pm}_1$   & $D$   & $g$   \\ \hline
-0.232     & -0.340     & -0.169       & -0.233       &  2.355  &  3.74  \\ \hline
\end{tabular}\\
\caption{Hamiltonian parameters that have been modified relative to
Ref.~\onlinecite{bai_hybridized_2020} to account for renormalization effects induced by the interactions between modes.}
\label{tab:1}
\end{table}

\begin{table}[h]
\centering
\begin{tabular}{c|c|c|c|c|c|c|c|c|c|c}
\hline
$J^{\pm}_2$ (meV)  & $J^{zz}_2$   & $J^{\pm}_3$   & $J^{zz}_3$   & $J^{'\pm}_0$     & $J^{'zz}_0$   & $J^{'\pm}_1$ & $J^{'zz}_1$  & $J^{'\pm}_{2a}$ & $J^{'zz}_{2a}$   \\ \hline
0.026     &   0.113     & 0.166      &  0.211    &  0.037  &  -0.036  &  0.013  &  0.051  &  0.068  &  0.073  \\ \hline
\end{tabular} \\
\caption{Spin-exchange tensors (beyond the nearest-neighbor bond) adopted in Ref.~\onlinecite{bai_hybridized_2020}.}
\label{tab:2}
\end{table}

As explained in the main text, the interaction effects captured by the above approach renormalize the energy spectrum of the low-energy modes. Consequently, the set of Hamiltonian parameters that reproduce the measured spectrum is not exactly the same as the one obtained with the {\it linear} generalized spin wave approach.~\cite{bai_hybridized_2020}
Table~\ref{tab:1} lists the Hamiltonian parameters that have been changed relative to the ones obtained in Ref.~\onlinecite{bai_hybridized_2020}, which includes the nearest-neighbor intra-plane spin-exchange tensor ${\cal J}_1$ (only $J_1^{zz}$ is actually different from Ref.~\onlinecite{bai_hybridized_2020}), the single-ion anisotropy $D$, and the $g$-factor along the $c$-axis.
The rest of the  Hamiltonian parameters, listed in Table~\ref{tab:2}, coincide with the ones adopted in Ref.~\onlinecite{bai_hybridized_2020}.

\section{Discussion of relation to older work}

\begin{table}[h!]
\centering
\begin{tabular}{ c|c|c }
\hline
Label & $H$ & $g_{eff}$ \\ \hline
1 & 0.6-3.5T & +4.6$\pm$0.1 \\ \hline
\multirow{2}{*}{2} & 1-1.5T & -4.7$\pm$0.3 \\
& 2.2-3.9T & -3.5$\pm$0.2 \\ \hline
\multirow{2}{*}{3} & 0-1.2T & +5.9$\pm$0.1 \\
& 2.2-3.5T & +3.5$\pm$0.1 \\ \hline
4 & 0-4.2T & -7.0$\pm$0.1  \\ \hline
\multirow{2}{*}{5} & 0.6-2T & -13.2$\pm$0.3 \\
& 2.7-3.7T & -12.6$\pm$0.4 \\ \hline
\multirow{2}{*}{6} & 0.2-1.8T & -14.4$\pm$0.2 \\
& 2.2-3.3T & -11.7$\pm$0.4 \\ \hline
\multirow{2}{*}{7} & 1.1-1.8T & -12.3$\pm$1.2 \\
& 2.2-3.4T & -14.9$\pm$0.2  \\ \hline
8 & 2.5-3.4T & -20.0$\pm$0.2  \\ \hline
9 & 3-3.5T & -20.7$\pm$0.4  \\ \hline
\end{tabular} \\
\caption{Estimated values of the slopes (in units of effective \textit{g}-factor) for the different excitation branches $\nu(H)$. A linear fit is performed on the field range where the field dependence is indeed linear (field range indicated in the table). Values can also change after a crossing between branches so there are sometimes several values for one branch, corresponding to different field ranges. The labels correspond to the ones used in Fig.3a in the main text.}
\label{tab:3}
\end{table}

The detection of many new resonances in comparison with previous work is obtained thanks to the fine tuning of magnetic field and the high frequency resolution achieved in these measurements.   The resonances that were previously observed (but without any differentiation of the two polarization channels RCP and LCP) are: the four branches starting at zero field, labeled 1 to 4 in Fig.3a \cite{petitgrand_magnetic_1980}, along with a few data points of branches in the region $\nu$ = 0.5-0.7 THz and \textit{H} = 2-4 T \cite{katsumata_single-ion_2000}. In these works, the character of the excitations were assigned based on their slopes (effective \textit{g}-factors).  They attributed branches 1 and 2, starting at $\nu$ $>$ 0.8 THz, to elementary single-magnon excitations (but they found two doublets instead of one).  The two other branches 3 and 4, starting at \textit{H} = 0 at lower energy ($\sim$0.65 THz), were assigned to single-ion two-magnon bound states; and the other branches in the region $\nu$ = 0.5-0.7 THz and \textit{H} = 2-4 T to a bound state of two single-ion bound states \textit{e.g.} a four-magnon bound state.

\begin{figure*}[h!]
\includegraphics[width=0.6\textwidth]{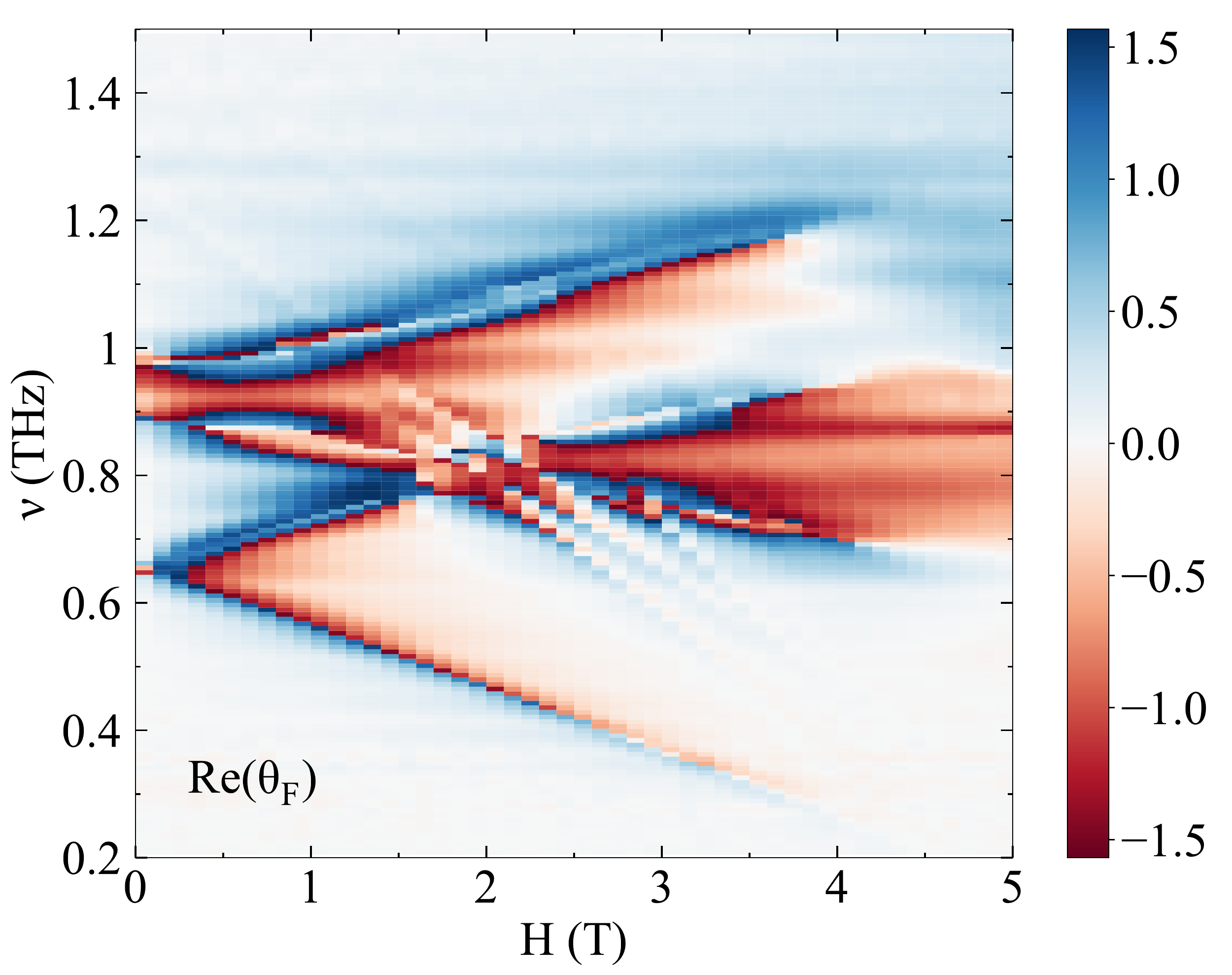}
\caption{Real part of the Faraday angle as a function of frequency and magnetic field at T = 4K.}%
\label{Fig.SI_1}%
\end{figure*}

The differences in the number of excitations for branches 1 and 2 (two doublets found in infrared spectroscopy instead of one) could be due to the fact that the previous infrared measurements were taken using a linear polarization basis, whereas we are displaying the transmission in the two circular polarization channels.  Indeed if the light is strongly rotated by the sample (the real part of the Faraday angle, plotted as a function of frequency and magnetic field in Fig.~\ref{Fig.SI_1}, lies between $\pm$ 1.5$^\circ$), then it may be misinterpreated as an absorption in such experiment and attributed to a magnetic excitation. In the presence of a magnetic field and in Faraday geometry, $T_R$ and $T_L$ are more suitable to study the magnetic excitations of the system. See for example the work by Pan \textit{et al.} \cite{pan2014low} (and the associated Supplementary Information) for a comparison of transmission data using linear and circular basis, and how the linear basis $T_{xx}$ and $T_{xy}$ cannot properly account for the positions of all magnetic excitations.

In the present work, the fine frequency steps of magnetic field and the high frequency resolution highlighted distinct new features such as excitation branches with an even higher slope in magnetic field than one of largely 4-magnon character, or a non-linear field dependence of excitations in some regions of the spectrum (such as branches 1 and 2 at low magnetic field, which show a quadratic field dependence for \textit{H} $<$ 1T, in contrast from what was inferred from previous infrared measurements). Finally, the theoretical calculations reported here were essential to confirm the strong hybridization between different \textit{n}-magnon states and to provide quantitative information on the mixing of different $\vert\Delta S^z\vert$ sectors. In this new scope, we understand that this lack of information on hybridization may led to misattribution of the excitation character in older infrared and ESR experiments. For additional information, we display in Table~\ref{tab:3} the values of the various slopes (in units of effective \textit{g}-factor) of the excitation branches observed in Fig.~3\textbf{a} of the main text. We use linear fits over field ranges where the field dependence of the absorption's position is indeed linear.

\newpage
\section{Temperature dependence of the terahertz spectra}

\begin{figure*}[h!]
\includegraphics[width=0.6\textwidth]{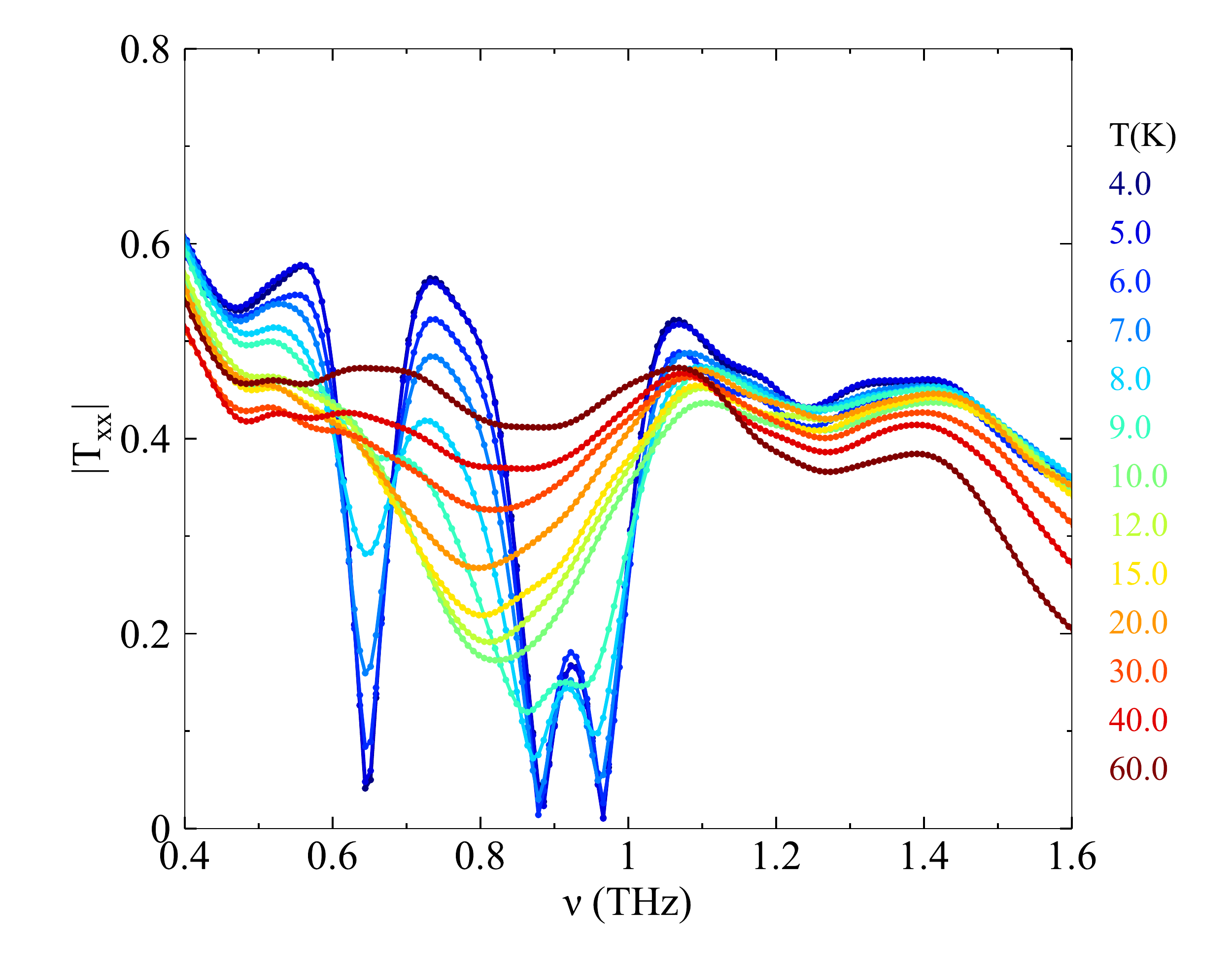}
\caption{Time-domain THz spectra of another FeI$_2$ crystal prepared in the same fashion at various temperatures in zero magnetic field. The time scans are short enough to avoid the Fabry-Perot effect discussed in the Methods.}%
\label{Fig.SI_2}%
\end{figure*}

\bibliographystyle{unsrtnat}
\bibliography{library}